\documentclass[sigconf]{acmart}
\AtBeginDocument{%
  }

\copyrightyear{2026}
\acmYear{2026}
\setcopyright{cc}
\setcctype{by-nc-nd}
\acmConference[MM '26]{Proceedings of the 34th ACM International Conference on Multimedia}{November 10--14, 2026}{Rio de Janeiro, Brazil}
\acmBooktitle{Proceedings of the 34th ACM International Conference on Multimedia (MM '26), November 10--14, 2026, Rio de Janeiro, Brazil}
\acmDOI{10.1145/3767308.3835759}
\acmISBN{979-8-4007-2213-4/2026/11}

\usepackage{pifont}
\usepackage{algorithm}
\usepackage{algorithmic}
\usepackage{amsmath}
\usepackage{amsfonts}
\usepackage{footmisc}
\usepackage{multirow}
\usepackage{booktabs}
\usepackage[utf8]{inputenc}
\usepackage{url}
\usepackage{bbding}
\usepackage{wasysym}
\usepackage{utfsym}
\usepackage{xcolor,soul}
\usepackage{fontawesome}
\usepackage{hyperref}
\usepackage{array}
\usepackage{svg}
\usepackage{tikz}
\usepackage{circledsteps}
\usepackage{colortbl}
\usepackage[most]{tcolorbox}
\usepackage{lipsum}
\usepackage{enumitem}
\usepackage{hhline}
\usepackage{multirow}
\usepackage[table,xcdraw]{xcolor}
\usepackage{makecell}
\newcommand{\ci}[1]{{\fontsize{6pt}{7pt}\selectfont\textcolor{gray}{$\pm$#1}}}
\usepackage{float}  
\begin{document}

\title{Let Me Look at You: Advanced Facial Expression Modeling for Conversational Speech Synthesis}

\author{Yifan Hu}
\email{32109235@mail.imu.edu.cn}
\orcid{0009-0008-2276-1456}
\affiliation{
  \institution{Inner Mongolia University}
  \city{Hohhot}
  \country{China}
}

\author{Shuwei He}
\email{shuwei\_he@163.com}
\orcid{0009-0003-2359-3523}
\affiliation{
  \institution{Inner Mongolia University}
  \city{Hohhot}
  \country{China}
}

\author{Rui Liu}
\email{imucslr@imu.edu.cn}
\correspondingauthor
\affiliation{
  \institution{Inner Mongolia University}
  \city{Hohhot}
  \country{China}
}

\author{Haizhou Li}
\email{haizhouli@cuhk.edu.cn}
\orcid{0000-0001-9158-9401}
\affiliation{
  \department{SRIBD, School of Artificial Intelligence}
  \institution{The Chinese University of Hong Kong}
 \city{Shenzhen}
 \country{China}
}


\renewcommand{\shortauthors}{Yifan Hu, Shuwei He, Rui Liu, and Haizhou Li.}

\begin{abstract}
Conversational Speech Synthesis~(CSS) is a fundamental component of human-computer interaction, aiming to generate contextually appropriate, expressive, and empathetic speech. 
However, facial expressions encode subtle and rich affective cues that are crucial for empathetic speech interaction, whereas existing approaches often overlook this important modality. 
In addition, the lack of large-scale natural conversational datasets with both speech and visual modalities also limits the development of visual affect understanding in conversational settings.
To address these limitations, we propose FacialTalker, a facial-expression-aware CSS framework built upon a large language model backbone. 
To efficiently encode facial expressions, we propose AUTokenizer, a single-codebook visual tokenizer that discretizes each frame-level facial expression into a compact token, trained with supervision from combinations of facial Action Units.
We further introduce a dual direct preference optimization (DualDPO) strategy, which extends the DPO by jointly imposing preference constraints on both visual and speech token sequences, to enhance the model's understanding of facial expressions and speech semantics in multimodal conversational contexts.
Moreover, we construct VSDD-1K, a large-scale multimodal dialogue dataset collected through a fully automated pipeline from real-world Internet conversations, comprising over 1,033 hours of synchronized speaker videos and speech, with more than 85\% of frames containing valid faces.
Extensive objective and subjective experiments demonstrate that FacialTalker consistently outperforms strong baselines in facial-expression perception and speech synthesis quality, generating speech that is more natural, expressive, and better aligned with the conversational context. The results also validate the effectiveness of our training strategy and dataset construction pipeline. The project page: https://github.com/walker-hyf/FacialTalker.
\end{abstract}

%
%

\ccsdesc[500]{Information systems~Multimedia content creation}

\keywords{Conversational Speech Synthesis, User-agent Conversation, Facial Expression Modeling, Empathetic Speech}


\maketitle

\begin{figure}[t]
\centering
\centerline{
\includegraphics[width=1\linewidth]{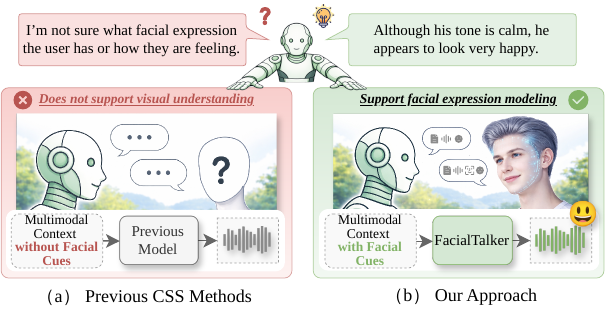}
}
\caption{(a) Previous CSS methods lack visual perception ability. (b) Our approach enables perception and modeling of the user’s facial expressions.}
\label{fig:demo}
\end{figure}

\section{Introduction}
Conversational speech synthesis~(CSS) aims to generate empathetic speech for the current user–agent interaction based on multimodal dialogue context~\cite{GRU-CSS}. Specifically, the synthesized speech should be not only natural, but also appropriate in speaking style and emotional expression~\cite{ECSS}. Despite recent advances in speech synthesis~\cite{vall-e2, indextts2, qwen3-tts, voxcpm}, existing methods still struggle to provide sufficient emotional support in complex multi-turn dialogues. This limitation is especially pronounced in practical applications such as smart homes~\cite{chatterjee2021real}, Smart Cockpit~\cite{wu2022toward}, and eldercare robots~\cite{wang2016user}, where users increasingly expect spoken interactions to convey not only information, but also richer emotional engagement.

Previous studies on CSS have improved speech naturalness and expressiveness through various context modeling approaches. By exploiting multi-scale information from text and speech in dialogue history at the sentence and word levels, ~\cite{M2-ctts, fctalker} introduce both coarse-grained and fine-grained contextual representations into the synthesizer to enhance contextual understanding. However, such methods often fail to capture complex contextual dependencies. To address this limitation, ~\cite{MSRGCN, ECSS} employ graph neural networks to model dependencies by representing multimodal information in each utterance as nodes and speaker relations and cross-modal interactions as edges, significantly improving contextual understanding. More recently, inspired by the autoregressive generation paradigm of large language models (LLMs), several studies~\cite{GPT-Talker, MultiDialog} further improve speech naturalness and emotional understanding with larger-scale training data. In parallel, LLM-based speech generation and spoken dialogue modeling have made remarkable progress in scenarios such as natural speech generation and spoken interaction~\cite{Speechgpt, E-chat, ParalinGPT, Step-audio2, Moshi}, demonstrating the strong potential of token-based unified modeling for text and speech. Nevertheless, most existing methods still primarily rely on textual and acoustic context, and thus remain limited in their ability to perceive users’ nonverbal emotional signals.

However, real-world interpersonal communication relies not only on spoken language for information exchange and emotion expression. Facial expressions, as a crucial source of nonverbal cues, play an equally important role in conveying affective states~\cite{bavelas2022face, leng2025eijl}. In particular, subtle micro-expressive changes in local facial regions often provide rich evidence for emotion perception~\cite{zhang2024review}. 
As shown in Fig.~\ref{fig:demo}, this modality remains largely underexplored in CSS due to two key challenges: (1) the lack of high-quality multimodal dialogue datasets with synchronized visual and speech signals~\cite{MultiDialog, IEMOCAP}, and (2) the limited ability of existing visual encoders to capture subtle facial expressions in conversational contexts~\cite{UniTalker, Empatheia}.
Therefore, how can users’ facial expressions be effectively perceived and encoded during contextual modeling for empathetic speech synthesis?

In this paper, we propose~\textbf{FacialTalker}, a novel CSS system with facial expression modeling. FacialTalker represents multimodal dialogue context, including speaker identity, text, speech, and face, as a unified sequence of discrete tokens for an LLM. Under the next-token prediction paradigm, the model generates the emotion category and speech token sequence for the target utterance, which are then rendered into expressive empathetic speech. 
To better capture facial affect, we develop \textbf{AUTokenizer}, a dedicated visual tokenizer that discretizes facial expressions into compact tokens under supervision from combinations of facial Action Units (AUs). AUTokenizer is highly efficient, requiring only a single token per frame, and is well suited for LLM-based autoregressive modeling.
We further propose a dual direct preference optimization strategy (\textbf{DualDPO}) that extends DPO to jointly impose preference constraints on visual and speech token sequences, encouraging the LLM to attend to both facial expressions and speech semantics within multimodal context modeling.
To address data scarcity, we construct \textbf{VSDD-1K}, a large-scale visual-speech dialogue dataset built through a fully automatic pipeline, comprising approximately 1,033 hours of data with comprehensive annotations, of which more than 85\% of the video frames contain valid faces.
Extensive objective and subjective evaluations demonstrate the effectiveness of VSDD-1K, the reliability of AUTokenizer, and the superiority of FacialTalker for empathetic speech synthesis. 

In summary, the main contributions of this paper are as follows:
\begin{itemize}
\item \textbf{Model}: We propose FacialTalker, a novel CSS model that incorporates facial affect modeling. We further develop AUTokenizer, an efficient visual tokenizer that encodes facial expression information into a single token per frame. 
\item \textbf{Training strategy}: We introduce a dual direct preference optimization strategy during post-training to further enhance the model’s ability to jointly model facial expressions and speech semantics.
\item \textbf{Dataset}: We develop a fully automatic pipeline for video-speech dialogue processing and construct VSDD-1K, a large-scale open-source video-speech dialogue dataset comprising approximately 1K hours of speech with well-aligned facial images for multimodal modeling.
\item  \textbf{Experiments}: Extensive objective and subjective experiments demonstrate the reliability and effectiveness of our model, training strategy, and dataset.
\end{itemize}

\section{Related Work}

\begin{table}[t]
\caption{\label{tab:dataset-1}VSDD-1K vs. other conversational datasets. $v$ $s$ and $t$ denote the visual, speech, and text modalities, respectively.}
\resizebox{1\linewidth}{!}{
\begin{tabular}{c|c|c|c|c}
\hline
\textbf{Dataset} & \textbf{Language} & \textbf{Modal} & \textbf{Duration (h)} & \textbf{Speaker} \\ \hline
NCSSD~\cite{GPT-Talker}  & CN, EN & ( ,a,t) & 236 & >776\\
RAMC~\cite{RAMC} & CN & (\textit{ ,a,t}) & 180 & 663 \\
DailyTalk~\cite{Dailytalk} & EN & (\textit{ ,a,t}) & 20 & 2 \\
STUDIES~\cite{saito2022studies} & JP & (\textit{ ,a,t}) & 8.2 & 3 \\
RyanSpeech~\cite{RyanSpeech} & EN & (\textit{ ,a,t}) & 10 & 1 \\
CALLS~\cite{saito2023calls} & JP & (\textit{ ,a,t}) & 6.5 & 1 \\ \hline
ECC~\cite{li2022ECC} & EN & (\textit{v,a,t}) & 24 & > 26 \\
MultiDialog~\cite{MultiDialog} & EN & (\textit{v,a,t}) & 340 & 12 \\
AvaMERG~\cite{Empatheia} & EN & (\textit{v,a,t}) & 239 & 65 \\ \hline
\rowcolor[HTML]{F5F9FE}
\textbf{VSDD-1K} & \textbf{EN} & \textbf{(\textit{v,a,t})} & \textbf{1033} & \textbf{\textgreater{}1498} \\ \hline
\end{tabular}
}
\end{table}

\begin{figure*}[t]
\centering
\centerline{
\includegraphics[width=1\linewidth]{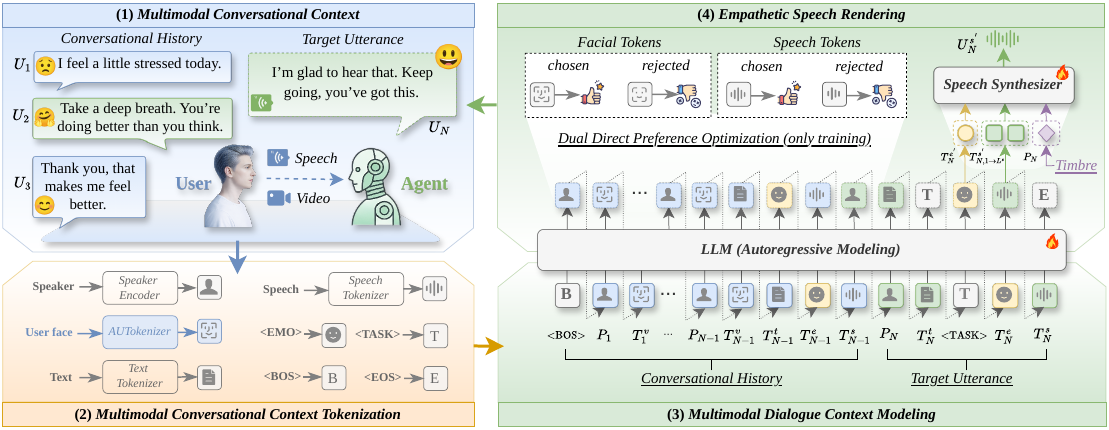}
}
\caption{The overview of \textbf{FacialTalker}. The framework consists of four main parts: (1) \textit{Multimodal Conversational Context} in the user-agent interaction, (2) multimodal information ($\mathcal{H}, \mathcal{C}$) tokenization via \textit{Multimodal Conversational Context Tokenization}, (3) feeding the serialized dialogue sequence into \textit{Multimodal Dialogue Context Modeling} to predict emotion categorie~$T_N^{e'}$ and speech tokens~$T_{N,1\rightarrow L}^{s'}$, with \textit{Dual Direct Preference Optimization} further enhancing visual and speech perception and modeling during training, and (4) \textit{Empathetic Speech Rendering} ($U_N^{s'}$) for the user.}
\label{fig:model}
\end{figure*}



\subsection{Multimodal Dialogue Dataset with Visual Signal}
Open-source multimodal multi-party dialogue datasets provide an important foundation for research on dialogue understanding and generation. For example, IEMOCAP~\cite{IEMOCAP}, MELD~\cite{MELD}, CPED~\cite{CPED}, and M$^3$ED~\cite{M3ED} support tasks such as conversational emotion recognition and dialogue act classification. However, their speech data often contain substantial noise and exhibit relatively low audio quality, which limits their value for training high-quality speech synthesis models.
To the best of our knowledge, Table~\ref{tab:dataset-1} summarizes currently available representative open-source datasets for target dialogue speech synthesis. 
NCSSD~\cite{GPT-Talker} uses a combination of data collection and manual recording to construct 236 hours of conversational speech synthesis data, including 188 hours in the collected subset and 48 hours in the recorded subset.
RAMC~\cite{RAMC} contains about 180 hours of speech manually recorded with smartphones in indoor environments. DailyTalk~\cite{Dailytalk} builds on DailyDialog~\cite{Dailydialog} scripts and provides about 20 hours of speech recorded by one male speaker and one female speaker. STUDIES~\cite{saito2022studies} records teacher-student interactions and contains 8.2 hours of speech. RyanSpeech~\cite{RyanSpeech} and CALLS~\cite{saito2023calls} contain 10 and 6.5 hours of conversational speech, respectively, but only preserve speech from one participant. Moreover, these datasets do not include visual signals and only provide speech and transcripts.
ECC~\cite{li2022ECC} collects about 66 English dialogue videos from YouTube, but it does not further process the videos for tasks such as target speaker extraction. MultiDialog~\cite{MultiDialog} and AvaMERG~\cite{Empatheia} provide manually recorded dialogue videos, but their speakers talk to the camera rather than engage in direct face-to-face interaction, which makes the conversations less realistic and natural. In contrast, VSDD-1K consists entirely of real-world face-to-face conversations collected from scenarios such as interviews and podcasts. As a result, it better reflects natural facial expressions during conversations.

\subsection{Facial Encoding in Multimodal Dialogue Modeling}
Visual encoders play a central role in computer vision because they directly affect how well a model understands visual information~\cite{pang2024mmaf}. Although recent spoken dialogue systems~\cite{team2025longcat, Qwen3-omni} and omni-modal models~\cite{Mini-omni2, ai2025ming} incorporate visual input, they mainly focus on scenes, objects, and other general visual content rather than user facial expressions.
Some spoken dialogue models begin to consider facial information. For example, Empatheia~\cite{Empatheia}, EmpathyEar~\cite{EmpathyEar}, and Mini-Omni2~\cite{Mini-omni2} use visual encoders such as CLIP~\cite{CLIP} for facial expression perception. However, these general-purpose pretrained models mainly optimize high-dimensional image-text alignment, which often leads to insufficient modeling of local facial details and unnecessary attention to non-facial regions. UniTalker~\cite{UniTalker} further explores facial expression modeling through facial landmark encoding. However, it uses only 128 landmarks and covers only the mouth, eyes, and eyebrows, which limits its ability to capture muscle movements from other facial regions.
Unlike previous work, we represent facial expressions using facial Action Units. We adopt AUs because individual units capture subtle local facial muscle movements, while combinations of multiple AUs provide a structured description of overall facial expressions~\cite{martinez2017automatic}. 
To better interface with LLMs, we design AUTokenizer to quantize facial expression information into discrete tokens, trained with supervision from AU combination labels. In contrast to general visual tokenizers~\cite{TamingVQGAN, Omnitokenizer}, which often require at least \(64 \times 64\) tokens to reconstruct an image or video frame, AUTokenizer leverages FSQ to quantize each facial frame into a single token while preserving rich facial expression information.

\begin{figure*}[t]
\centering
\centerline{
\includegraphics[width=1\linewidth]{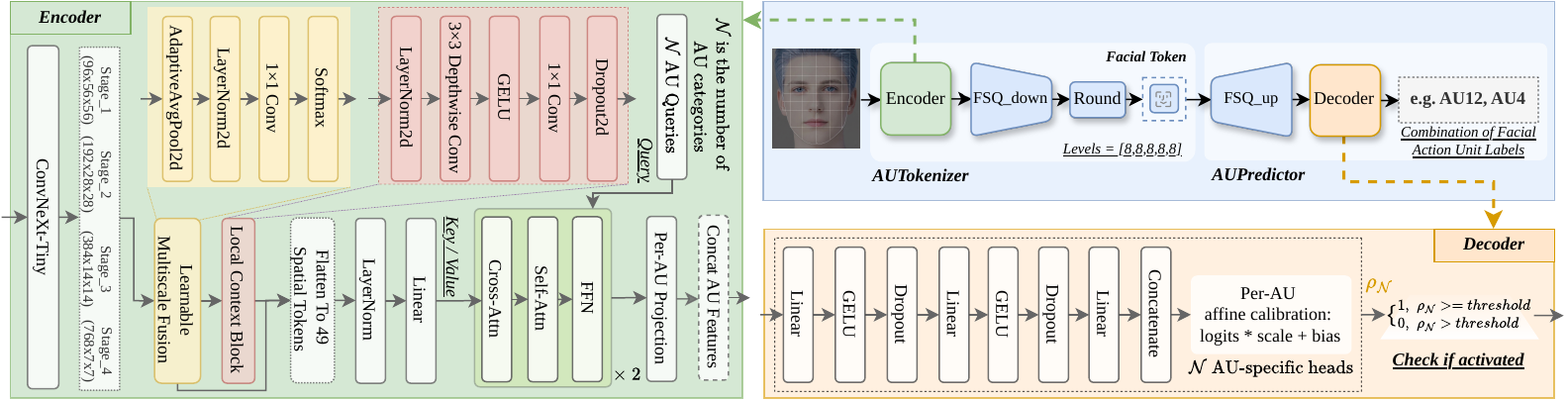}
}
\caption{The overall architecture of AUTokenizer.}
\label{fig:codec}
\end{figure*}

\section{Task Definition}


The goal is to enable the model to understand the user's facial expressions \(U_i^v\) and speech \(U_i^s\) through contextual modeling, and then generate an appropriate target emotion \(U_N^{e'}\) and target speech \(U_N^{s'}\). In particular, \(U_N^{s'}\) should not only natural and remain consistent with \(U_N^{e'}\), but also match the style, timbre, and emotion of the overall dialogue context.
In CSS, the dialogue context typically consists of two parts: the dialogue history \(\mathcal{H}\) and the target utterance \(\mathcal{C}\). The dialogue history \(\mathcal{H}\) contains \(N-1\) utterances, denoted as \(\{U_i\}_{i=1}^{N-1}\). Each utterance \(U_i\) is represented as a five-tuple \((U_i^p, U_i^v, U_i^t, U_i^e, U_i^s)\), where \(U_i^p\), \(U_i^v\), \(U_i^t\), \(U_i^e\), and \(U_i^s\) denote the speaker identity, facial expression, text, emotion label, and speech, respectively, and \(i\) denotes the dialogue turn. 
Notably, we only include the facial visual modality \(U_i^v\) for user turns, which better reflects real-world application scenarios.
The target utterance \(\mathcal{C}\) consists of the agent identity \(U_N^p\) and the target text \(U_N^t\) to be synthesized at turn \(N\). Given the multimodal dialogue history \(\mathcal{H}\) and the target utterance \(\mathcal{C}\), the model \(\mathcal{F}\) aims to predict the target emotion \(U_N^{e'}\) and the target speech \(U_N^{s'}\). The task is formulated as:
\begin{equation}
\begin{array}{c}
\{U_N^{e'}, U_N^{s'}\} = \mathcal{F}(\mathcal{H}, \mathcal{C}), \\
\mathcal{H} = \{(U^p_i, U^v_i, U^t_i, U^e_i, U^s_i)\}_{i=1}^{N-1}, \quad
\mathcal{C} = \{U^p_N, U^t_N\}
\end{array}
\end{equation}
where
\begin{equation}
U^p_i =
\begin{cases}
\text{None}, & \text{if } U^p_i = U^p_N \\
U^p_i, & \text{otherwise}
\end{cases}
\end{equation}

\section{FacialTalker: Methodology}
Fig.~\ref{fig:model} illustrates the overall architecture of FacialTalker, which consists of four components:
(1) \textit{Multimodal Conversational Context}, providing the multimodal context ($\mathcal{H}, \mathcal{C}$) between the user and the agent as model input.
(2) \textit{Multimodal Conversational Context Tokenization}, which extracts initial representations from raw inputs, including speaker embedding ($U^p_i$) and tokenization of multimodal signals ($U^v_i, U^t_i, U^e_i, U^s_i$).
(3) \textit{Multimodal Dialogue Context Modeling}, where the LLM serializes multimodal representations and autoregressively predicts the target emotion category and speech token sequence.
(4) \textit{Empathetic Speech Rendering}, where the Speech Synthesizer combines the agent timbre with predicted emotion and speech tokens to generate an empathetic response.

\subsection{Multimodal Conversational Context Tokenization}
Traditional CSS models usually depend on multiple pretrained models to extract representations at different scales and require extra modules for dimensional alignment. In contrast, our method directly adopts the classical LLM data processing paradigm and tokenizes raw data into discrete representations. For speaker identity modeling, using discrete labels directly limits generalization to out-of-distribution data. Therefore, we employ a pretrained model to derive continuous speaker representations.

\subsubsection{\textbf{Text, Emotion, and Speech Tokenizers}}
We use the classical Byte Pair Encoding (BPE)~\cite{BPE} tokenizer to discretize raw text $U_i^t$ into $T_i^t$. BPE relies on frequency-based subword merging and strikes a balance among vocabulary size, sequence length, and open-vocabulary coverage:
\[
T^t_{i,1 \rightarrow L^t} = \text{TextTokenizer}(U_i^t)
\]
where $L^t$ is the length of the text token sequence.

For the eight emotion categories ($\text{<EMO>}$), we use the same tokenization scheme as for text. For speech, we follow CosyVoice2~\cite{cosyvoice2} and insert an FSQ~\cite{FSQ} module after the encoder of the SenseVoice-Large~\cite{SenseVoice} ASR model to discretize raw speech $U_i^s$ into $T_i^s$:
\[
T^s_{i,1 \rightarrow L^s} = \text{SpeechTokenizer}(U_i^s)
\]

\subsubsection{\textbf{Speaker Encoder and Special Tokens}}
Speaker identity is directly associated with dialogue turns in context modeling. However, a single speaker label loses speaker-specific characteristics across different speakers. Therefore, our Speaker Encoder uses a pre-trained voiceprint model~\footnote{\label{cam++}https://github.com/modelscope/3D-Speaker/tree/main/egs/3dspeaker/svcam++} to extract a speaker representation $P_i$ from raw speech $U_i^s$:
\[
P_i = \text{SpeakerEncoder}(U_i^s)
\]

In addition, we introduce three special tokens in the context modeling process. $\text{<BOS>}$ and $\text{<EOS>}$ are the beginning and the end of the whole dialogue sequence, respectively, and $\text{<TASK>}$ denotes the task that the model needs to perform. In this work, starting from $\text{<TASK>}$, the LLM first predicts the emotion category and then autoregressively predicts speech tokens until it outputs $\text{<EOS>}$.

\subsubsection{\textbf{AUTokenizer}}
As shown in Fig.~\ref{fig:codec}, we propose a novel visual tokenizer, termed AUTokenizer, to compactly encode facial expressions. The model consists of three components: an encoder, an FSQ-based quantization module, and a decoder.

Given an input facial image $U^v_{i,j}$, we first employ a ConvNeXt-Tiny~\cite{liu2022convnet} backbone to extract hierarchical features $\{f_k\}_{k=1}^4$. To mitigate the discrepancy of facial information across scales, we fuse these features using a Learnable Multiscale Fusion Module:
\begin{equation}
F_{\text{fuse}} = \sum_{k=1}^{4} w_k \cdot f_k,
\end{equation}
where $w_k$ are learnable weights that adaptively balance facial information from coarse to fine scales. The fused features are further refined by a local context block and flattened into a sequence of spatial tokens $\mathcal{T}$.
To model different facial regions associated with expressions, we introduce $\mathcal{N}$ learnable AU queries $\mathcal{Q}_{\text{AU}}$, which focus on AU-relevant regions and extract discriminative expression features. A Transformer-based decoder captures the interaction between spatial features and AU queries via cross-attention:
\begin{equation}
\mathcal{Q}_{\text{AU}}^{\text{out}} = \text{Attention}(\mathcal{Q}_{\text{AU}}, \mathcal{T}, \mathcal{T}).
\end{equation}
This process injects global spatial context into AU queries, enabling the model to disentangle expressive motion from static facial identity.
The resulting AU features are projected and concatenated into a compact 128-dimensional representation. We then apply an FSQ-based quantization module, consisting of two linear layers (FSQ\_down) followed by a rounding operation:
\begin{equation}
T_{i,j}^v = \text{ROUND}(\text{FSQ\_down}(\text{Encoder}(U^v_{i,j}))).
\end{equation}
The encoder, FSQ\_down, and ROUND jointly constitute the \textbf{AUTokenizer}, which maps each facial frame into a single discrete token.

To recover semantic facial information, we introduce an \textbf{AUPredictor}, composed of three linear layers (FSQ\_up) and AU-specific decoding heads. The quantized token is first projected back to 128 dimensions and then decoded into AU activations:
\begin{equation}
\text{AUs}_{i,j} = \text{Decoder}(\text{FSQ\_up}(T_{i,j}^v)).
\end{equation}
By supervising the model with AU combination labels, AUTokenizer learns to encode rich facial expression information into compact discrete tokens.

\subsection{Multimodal Dialogue Context Modeling and Empathetic Speech Rendering}
We employ a LLM based on Qwen2.5-0.5B to model multimodal conversational context. The LLM takes a discretized multimodal token sequence as input and performs autoregressive next-token prediction. Because emotion directly relates to the expressiveness of the target utterance, we require the LLM to first predict an appropriate emotion category $T^{e'}_N$ for the target utterance and then generate the target speech token sequence $T^{s'}_{N,1 \rightarrow L^s}$ token by token.

For the speech rendering, we adopt emotion-guided Conditional Flow Matching (CFM)~\cite{CFM} to generate empathetic and expressive speech. Specifically, the Speech Synthesizer takes four conditions as input: the speech token sequence $T^{s'}_{N,1 \rightarrow L^s}$, the emotion $T^{e'}_N$, the agent identity $P_N$, and a reference mel-spectrogram extracted from the agent's speech in the most recent turn. Based on these conditions, a causal convolutional Transformer UNet serves as the vector field predictor and samples the mel-spectrogram of the target speech. We then use a HiFi-GAN vocoder~\cite{Hifi-gan} to convert the generated mel-spectrogram into a speech waveform for user response. Following CosyVoice2~\cite{cosyvoice2}, we employ an Optimal Transport flow to guide vector field learning in CFM training. The objective drives the predicted vector field toward the ground-truth one.

\subsection{Training Strategies}
To further improve each component of FacialTalker, we adopt a post-training strategy to enhance the language model’s understanding of facial expressions and speech semantics. Meanwhile, during AUTokenizer training, we optimize the model using a combination of loss functions to mitigate label imbalance in the dataset.

\subsubsection{\textbf{Training for LM}}
For the LLM, training is conducted in four stages. 
In \textbf{Stage-1}, we initialize the model from CosyVoice2, pretrained on 170k hours of high-quality single-utterance speech data. 
In \textbf{Stage-2}, we fine-tune the model on dialogue speech synthesis datasets to improve speech understanding. 
In \textbf{Stage-3}, we further fine-tune it on a video-speech dialogue dataset to enhance visual-speech understanding, using a cross-entropy loss over facial, speech, and emotion tokens against their ground-truth targets. 
In \textbf{Stage-4}, after training stabilizes, we apply a dual direct preference optimization strategy for post-training to further improve model stability and contextual understanding in multimodal settings.

Specifically, (1) for the speech modality, we re-tokenize the speech generated by the fine-tuned Stage-3 model using the Speech Tokenizer as the rejected sample, while the discretized ground-truth speech token sequence is used as the chosen sample. 
(2) For the visual modality, we decode and sample the facial tokens generated during autoregressive modeling by the fine-tuned Stage-3 model as the rejected sample, while the token sequence obtained by discretizing real facial data with AUTokenizer is used as the chosen sample. 
Both the policy and reference models are initialized from the fine-tuned Stage-3 model. The reference model remains frozen during training to facilitate reward loss computation and prevent the model from deviating excessively from the initial SFT model. 
During training, we jointly optimize the cross-entropy losses of the three modalities in the LLM together with the two DPO losses.
\subsubsection{\textbf{Training for AUTokenizer}}
For AUTokenizer, AU prediction datas\-ets usually exhibit class imbalance. Therefore, we optimize the model with AsymmetricLoss~\cite{AsymmetricLoss} and SoftMacroF1Loss~\cite{SoftMacroF1Loss}. This design mitigates class imbalance and improves learning of under-represented AU patterns.

\begin{figure}[t]
\centering
\centerline{
\includegraphics[width=1\linewidth]{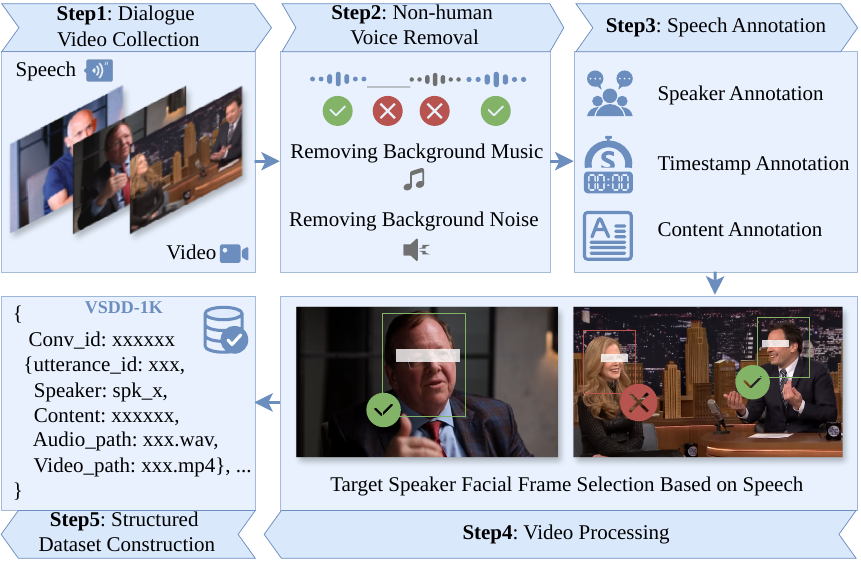}
}
\caption{The construction pipeline of the VSDD-1K visual-speech dialogue dataset.}
\label{fig:pipeline}
\end{figure}

\section{VSDD-1K: Dataset}
As mentioned earlier, we construct a large-scale visual–speech dialogue dataset, VSDD-1K. The dataset is released under the \textit{CC BY-NC-SA 4.0} license to support the research community across a broad range of multimodal dialogue tasks. 
In the following, we introduce the dataset from two aspects: (1) \textit{Automatic Dataset Construction Pipeline} and (2) \textit{Dataset Statistics}.

\subsection{Automated Dataset Construction Pipeline}
As shown in Fig.~\ref{fig:pipeline}, our pipeline consists of five stages.

\subsubsection{\textbf{Step-1: Dialogue Video Collection.}}
Unlike previous studies that primarily select dialogue videos from TV shows~\cite{li2022ECC, GPT-Talker}, we focus on videos from scenarios such as interviews and face-to-face podcasts. We make this choice for two main reasons. First, movies and television dramas usually contain more background music, which significantly degrades speech quality. Second, frequent scene transitions and variations between close-up and long shots often make the target speaker’s face difficult to capture reliably. In total, we collect 1,362 real-world dialogue videos.

\subsubsection{\textbf{Step-2: Non-human Voice Removal.}}
At this stage, we first use a voice activity detection (VAD) model~\footnote{https://github.com/snakers4/silero-vad} to remove segments with pauses longer than 5 seconds, because such pauses usually indicate the beginning of a new dialogue turn. We then apply a high-performance vocal-background separation tool~\footnote{https://github.com/facebookresearch/demucs} and compute the signal-to-noise ratios (SNR) of the vocal signal and the background noise. Based on these measurements, we identify and remove regions whose values fall below 4, thereby ensuring the cleanliness of the speech signal.

\subsubsection{\textbf{Step-3: Speech Annotation.}}
To obtain more accurate annotations of speakers, transcripts, and timestamps, we use Pyannote’s transcription-precision-2 API~\footnote{https://docs.pyannote.ai/tutorials/speech-to-text-diarization} for speech recognition.

\subsubsection{\textbf{Step-4: Video Processing.}}
This stage focuses on determining whether the speech signal matches the face that appears in the current video frame. To this end, we use a strong Active Speaker Detection (ASD) model~\cite{liao2025lrasd} to evaluate audio-visual consistency and save the facial regions that are consistent with the speech. For the filtered-out video frames, we use the special $\text{<IGNORE>}$ label during the training of FacialTalker. This strategy maintains alignment between the lengths of visual tokens and speech tokens.

\subsubsection{\textbf{Step-5: Structured Dataset Construction.}}
Finally, we standardize all videos to 25 FPS and all audio tracks to 16 kHz mono. As shown in Step 5 of Fig.~\ref{fig:pipeline}, we record the information for each dialogue pair in dictionary-style format.

\subsection{Data Statistics}
The VSDD-1K contains 61,801 dialogues and 991,439 utterances, covering a broad range of topics, including culture, sports, humanities, economics, and healthcare, with a total duration of approximately 1,033 hours. Each conversation contains between 3 and 20 utterances, with an average of 16.04, providing sufficiently rich conversational context for dialogue-related tasks. Conversation durations range from 5.00 to 353.23 seconds, with an average of 64.60 seconds. Each utterance contains an average of 12.59 words and has an average duration of 3.75 seconds, with durations ranging from 0.06 to 104.88 seconds. Moreover, at least 85\% of the video frames contain valid facial information from the target speaker, providing sufficient visual cues for facial expression understanding.

\section{Experiments}
In this section, we first describe the datasets used for training and evaluating AUTokenizer and FacialTalker, along with the baseline models and the objective and subjective evaluation metrics.

\subsection{Datasets}
We use two types of datasets for training.
(1) \textbf{AUTokenizer training}. We use CASME II~\cite{CASME-II} and DISFA~\cite{DISFA}, two datasets designed for micro-expression recognition. Each sample consists of a facial image and its manually annotated AU combination. CASME II contains 247 facial video samples from 26 subjects with annotations for eight AUs: 1, 2, 4, 7, 12, 14, 15, and 17. DISFA contains 27 facial video samples from 27 subjects with annotations for eight AUs: 1, 2, 4, 6, 9, 12, 25, and 26. We employ the Leave-One-Subject-Out cross-validation protocol~\cite{varanka2023learnable} to ensure subject-independent evaluation.
(2) \textbf{FacialTalker training}. We use two speech dialogue datasets (Datasets-S): DailyTalk~\cite{Dailytalk} and the English subset of NCSSD~\cite{GPT-Talker}, and three visual--speech dialogue datasets (Datasets-VS): VSDD-1K, MultiDialog~\cite{MultiDialog}, and AvaMERG~\cite{Empatheia}. The total duration of Datasets-S is 112 hours, while Datasets-VS total approximately 1,612 hours. All datasets are divided into training, validation, and test sets in an 8:1:1 ratio.

\subsection{Baseline Models}
To evaluate FacialTalker, we compare it with state-of-the-art models from three perspectives.
(1) \textbf{Facial action unit classification}. We evaluate AUTokenizer’s AU classification capability by comparing it with recognition models, including ME-GraphAU~\cite{ME-GraphAU}, VL-FAU~\cite{VL-FAU}, AU-LLaVA~\cite{AU-LLaVA}, and EmoLA~\cite{EmoLA}, which are fine-tuned on DISFA, as well as SCA~\cite{li2021SCA}, DVASP~\cite{li2021micro}, SSSNet-LED~\cite{varanka2023learnable}, and AULLM~\cite{liu2025aullm}, which are fine-tuned on CASME II.
(2) \textbf{Conversational speech synthesis without visual input}. We evaluate FacialTalker’s contextual modeling ability by comparing it with GRU-CSS~\cite{GRU-CSS}, a GRU-based method; M$^2$CTTS~\cite{M2-ctts}, a multi-scale modeling approach; MSRGCN~\cite{MSRGCN} and ECSS~\cite{ECSS}, graph neural network-based methods; and GPT-Talker~\cite{GPT-Talker}, a GPT-2-style decoder-only Transformer model.
(3) \textbf{Conversational speech synthesis with visual input}. We evaluate FacialTalker’s visual modeling capability by comparing it with Empatheia~\cite{Empatheia} and EmpathyEar~\cite{EmpathyEar}, which leverage pre-trained visual models~\cite{CLIP}, as well as UniTalker~\cite{UniTalker}, which is based on a landmark codec design.
FacialTalker has two variants: FacialTalker (D) and FacialTalker (C), which use AUTokenizer trained on DISFA and CASME II, respectively.
In addition, to evaluate the contribution of each component, we construct three variants: AUTokenizer-VQ, which replaces FSQ with a VQ layer to validate the quantization strategy; FT-base, which uses AUTokenizer (D) and removes the DualDPO post-training strategy; and FT-CLIP, which incorporates a CLIP-based visual encoder to examine AUTokenizer’s visual perception capability.

\subsection{Evaluation Metrics}
We evaluate each model using both objective and subjective metrics.
(1) \textbf{Objective metrics}. 
\textit{For AU classification}, we use the F1 score, the harmonic mean of precision and recall, where higher values indicate better performance. 
\textit{For speech synthesis}, we measure speaker similarity using WavLM~\cite{chen2022wavlm} and CAM++ scores~\footref{cam++} (\(\mathrm{SIM}_{\text{WavLM}}\) and \(\mathrm{SIM}_{\text{CAM}}\)). We assess prosody using the Dynamic Time Warping (DTW) distance between the pitch distributions of generated and real speech (PDTW)~\cite{ren2020fastspeech}, where higher similarity scores and lower PDTW indicate better performance. 
We use Emotion2vec~\cite{ma2024emotion2vec} to compute emotion classification accuracy (ACC\(_E\)) and evaluate emotional expression. We also use DNSMOS~\cite{reddy2021dnsmos} and UTMOS~\cite{saeki2022utmos} to evaluate overall speech quality.
\textit{For images in the training set}, we use CR-FIQA~\cite{CR-FIQA} and IFQA~\cite{jo2023ifqa} to assess overall facial image quality. 
(2) \textbf{Subjective metrics}. We recruit 50 participants who speak English as a second language and have good listening and speaking skills. After training, they complete the evaluation in a quiet environment with reference to the dialogue context. 
Each sample is rated on a 1–5 scale from poor to excellent. Two aspects are evaluated: MOS\(_N\), which measures whether the synthesized speech is natural and fits the dialogue context, and MOS\(_E\), which measures whether the expressed emotion is accurate and contextually appropriate.

\section{Results and Discussions}
In this section, we present a comprehensive evaluation of VSDD-1K, AUTokenizer and FacialTalker. 

\begin{table}[t]
\caption{\label{tab:dataset-3} Quality comparison of VSDD-1K and other datasets.}
\resizebox{1\linewidth}{!}{
\begin{tabular}{c|cccccc}
\hline
 & \multicolumn{2}{c|}{\textbf{Speech Quality}} & \multicolumn{2}{c|}{\textbf{Speaker Similarity}} & \multicolumn{2}{c}{\textbf{Image Quality}} \\ \cline{2-7} 
 & DNSMOS & \multicolumn{1}{c|}{UTMOS} & SIM$_{WavLM}$ & \multicolumn{1}{c|}{{SIM$_{CAM}$}} & CR-FIQA & IFQA \\ \hhline{~|------}
\multirow{-3}{*}{\textbf{Datasets}} & \multicolumn{6}{c}{\cellcolor[HTML]{EFEFEF}\textit{Higher values indicate better performance.} ($\uparrow$)} \\ \hhline{-|------}
MultiDialog~\cite{MultiDialog} & \textbf{3.12} & \multicolumn{1}{c|}{2.91} & 0.96$^*$ & \multicolumn{1}{c|}{0.89} & 4.43$^*$ & 0.42$^*$ \\
AvaMERG~\cite{Empatheia} & 2.92 & \multicolumn{1}{c|}{\textbf{3.11}} & \textbf{0.97} & \multicolumn{1}{c|}{\textbf{0.93}} & 2.70 & 0.10 \\ \hhline{-|------}
\rowcolor[HTML]{F5F9FE}
\textbf{VSDD-1K} & 3.10$^*$ & \multicolumn{1}{c|}{2.93$^*$} & \textbf{0.97} & \multicolumn{1}{c|}{0.92$^*$} & \textbf{4.57} & \textbf{0.45} \\ \hhline{-|------}
\end{tabular}
}
\end{table}

\subsection{VSDD-1K Reliability Verification}
Existing visual–speech dialogue datasets~\cite{MultiDialog, Empatheia} have demonstrated their effectiveness in supporting visual understanding and speech synthesis for contextual modeling in CSS. We therefore conduct a comparison from three perspectives. 
Specifically, we randomly sample 400 same-speaker utterance pairs from each dataset and compare them in terms of speech quality, speaker similarity, and facial image quality. DNSMOS and UTMOS results show that VSDD-1K achieves the second-best speech quality and remains close to the recorded speech in MultiDialog. VSDD-1K also performs well in terms of speaker similarity and facial image quality. 
Overall, these results indicate that our dataset effectively supports multimodal dialogue context understanding and speech generation.

\subsection{AUTokenizer Reliability Verification}

\begin{table}[t]
\caption{\label{tab:exp-AUTokenizer} Comparison of AU label classification performance between AUTokenizer and other baselines.}
\resizebox{1\linewidth}{!}{
\begin{tabular}{cccccccccc}
\hhline{----------}
\multicolumn{1}{c|}{\textbf{Methods}} & \multicolumn{9}{c}{\textbf{F1} $\uparrow$} \\ 
\hhline{----------}
\rowcolor[HTML]{EFEFEF} 
\multicolumn{10}{c}{\cellcolor[HTML]{EFEFEF}\textit{Dataset  DISFA}} \\ 
\hhline{----------}
\multicolumn{1}{c|}{-} & AU1 & AU2 & AU4 & AU6 & AU9 & AU12 & AU25 & \multicolumn{1}{c|}{AU26} & Avg. \\ 
\hhline{----------}
\multicolumn{1}{c|}{ME-GraphAU~\cite{ME-GraphAU}} & 0.55 & 0.47 & 0.73 & 0.54$^*$ & 0.56 & 0.77$^*$ & 0.91 & \multicolumn{1}{c|}{0.53} & 0.63 \\
\multicolumn{1}{c|}{VL-FAU~\cite{VL-FAU}} & 0.61$^*$ & 0.56 & 0.74 & 0.46 & \textbf{0.61} & 0.72 & \textbf{0.94} & \multicolumn{1}{c|}{\textbf{0.67}} & \textbf{0.66} \\
\multicolumn{1}{c|}{AU-LLaVA~\cite{AU-LLaVA}} & 0.52 & 0.59 & 0.44 & 0.31 & 0.22 & 0.66 & 0.91 & \multicolumn{1}{c|}{0.55} & 0.53 \\
\multicolumn{1}{c|}{EmoLA~\cite{EmoLA}} & 0.51 & 0.57 & 0.84$^*$ & \textbf{0.55} & 0.43 & \textbf{0.80} & 0.92$^*$ & \multicolumn{1}{c|}{0.60$^*$} & 0.65$^*$ \\ 
\multicolumn{1}{c|}{AUTokenizer-VQ} & 0.52 & 0.61$^*$ & 0.66 & 0.39 & 0.52 & 0.49 & 0.75 & \multicolumn{1}{c|}{0.44} & 0.55 \\ \hline
\rowcolor[HTML]{F5F9FE} 
\multicolumn{1}{c|}{\textbf{AUTokenizer~(D)}} & \textbf{0.69} & \textbf{0.73} & \textbf{0.91} & 0.38 & 0.60$^*$ & 0.58 & 0.84 & \multicolumn{1}{c|}{0.46} & 0.65$^*$ \\ 
\hhline{----------}
\rowcolor[HTML]{EFEFEF} 
\multicolumn{10}{c}{\textit{Dataset  CASME II}} \\ 
\hhline{----------}
\multicolumn{1}{c|}{-} & AU1 & AU2 & AU4 & AU7 & AU12 & AU14 & AU15 & \multicolumn{1}{c|}{AU17} & Avg. \\ 
\hhline{----------}
\multicolumn{1}{c|}{SCA~\cite{li2021SCA}} & 0.64 & 0.77 & 0.81 & 0.56 & 0.61 & 0.62 & 0.69 & \multicolumn{1}{c|}{0.64} & 0.67 \\
\multicolumn{1}{c|}{DVASP~\cite{li2021micro}} & 0.73 & 0.72 & 0.90$^*$ & 0.57 & 0.80 & 0.69 & 0.72 & \multicolumn{1}{c|}{0.70} & 0.73 \\
\multicolumn{1}{c|}{SSSNet-LED~\cite{varanka2023learnable}} & \textbf{0.93} & 0.84$^*$ & 0.89 & 0.64$^*$ & 0.77$^*$ & 0.74$^*$ & 0.72 & \multicolumn{1}{c|}{0.76$^*$} & 0.79$^*$ \\
\multicolumn{1}{c|}{AULLM~\cite{liu2025aullm}} & 0.92$^*$ & \textbf{0.87} & 0.90$^*$ & \textbf{0.68} & \textbf{0.79} & \textbf{0.75} & 0.79$^*$ & \multicolumn{1}{c|}{\textbf{0.81}} & \textbf{0.81} \\ 
\multicolumn{1}{c|}{AUTokenizer-VQ} & 0.58 & 0.71 & 0.86 & 0.59 & 0.67 & 0.70 & 0.74 & \multicolumn{1}{c|}{0.71} & 0.70 \\ \hline
\rowcolor[HTML]{F5F9FE} 
\multicolumn{1}{c|}{\textbf{AUTokenizer~(C)}} & 0.69 & 0.78 & \textbf{0.91} & 0.62 & 0.72 & 0.71 & \textbf{0.82} & \multicolumn{1}{c|}{0.73} & 0.75 \\ 
\hhline{----------}
\end{tabular}
}
\end{table}

We evaluate our method on DISFA and CASME II. As shown in Table~\ref{tab:exp-AUTokenizer}, compared with prior methods that either model continuous facial representations or prompt LMs to predict AU categories, AUTokenizer achieves the second-best performance on DISFA with an average F1 score of 0.65, and also demonstrates competitive performance on CASME II with an average F1 score of 0.75. 
Compared with AUTokenizer-VQ, it improves the average F1 score by approximately 0.10 and 0.05 on DISFA and CASME II, respectively, further validating the effectiveness of FSQ-based quantization. 
Overall, these results demonstrate that our method effectively encodes facial information into a single token, providing a strong foundation for subsequent LLM-based contextual modeling.

\begin{table*}[t]
\caption{\label{tab:exp-facialtalker-1} Subjective (95\% confidence interval) and objective experimental results on MultiDialog and AvaMERG datasets.}
\vspace{-2mm}
\resizebox{0.9\linewidth}{!}{
\begin{tabular}{l|ccc|cc|ccc|cc}
\hhline{-|---|--|---|--}
\multirow{2}{*}{\textbf{Models}}
    & \makecell{\textbf{SIM}$\uparrow$}
  & \makecell{\textbf{PDTW}$\downarrow$}
  & \makecell{\textbf{ACC$_{E}$}$\uparrow$}
  & \multicolumn{1}{c}{\makecell{\textbf{MOS$_{N}$} $\uparrow$}}
  & \multicolumn{1}{c|}{\makecell{\textbf{MOS$_{E}$} $\uparrow$}}
    & \makecell{\textbf{SIM}$\uparrow$}
  & \makecell{\textbf{PDTW}$\downarrow$}
  & \makecell{\textbf{ACC$_{E}$}$\uparrow$}
  & \multicolumn{1}{c}{\makecell{\textbf{MOS$_{N}$}$\uparrow$}}
  & \multicolumn{1}{c}{\makecell{\textbf{MOS$_{E}$}$\uparrow$}} \\
\cline{2-11}
\hhline{~|---|--|---|--}
  & \multicolumn{5}{c|}{\cellcolor[HTML]{EFEFEF} \textit{Dataset MultiDialog}} 
  & \multicolumn{5}{c}{\cellcolor[HTML]{EFEFEF}  \textit{Dataset AvaMERG}} \\ 
\hhline{-|---|--|---|--}
\rowcolor[HTML]{FDF0EF} 
\multicolumn{1}{l|}{Ground Truth} & - & - & - & 4.52\ci{0.03} & 4.32\ci{0.12} & - & - & - & 4.31\ci{0.04} & 4.35\ci{0.02} \\
\hhline{-|---|--|---|--}
GRU-CSS~\cite{GRU-CSS}    & 0.73 & 68.21 & 0.55 & 3.40\ci{0.01} & 3.45\ci{0.02} & 0.68 & 73.65 & 0.49 & 3.51\ci{0.02} & 3.56\ci{0.02} \\
M$^2$CTTS~\cite{M2-ctts}  & 0.75 & 67.02 & 0.63 & 3.62\ci{0.02} & 3.60\ci{0.03} & 0.71 & 72.10 & 0.56 & 3.60\ci{0.04} & 3.59\ci{0.01} \\
MSRGCN~\cite{MSRGCN}     & 0.75 & 64.25 & 0.63 & 3.64\ci{0.03} & 3.65\ci{0.02} & 0.70 & 70.42 & 0.58 & 3.62\ci{0.03} & 3.64\ci{0.02} \\
ECSS~\cite{ECSS}       & 0.76 & 58.16 & 0.66 & 3.76\ci{0.02} & 3.76\ci{0.03} & 0.73 & 66.13 & 0.61 & 3.69\ci{0.01} & 3.68\ci{0.01} \\
GPT-Talker~\cite{GPT-Talker} & 0.87 & 47.27 & 0.68 & 3.91\ci{0.02} & 3.91\ci{0.02} & 0.85 & 50.16 & 0.62 & 3.86\ci{0.03} & 3.85\ci{0.04} \\ 
\hhline{-|---|--|---|--}
Empatheia~\cite{Empatheia}  & 0.82 & 50.49 & 0.71 & 3.91\ci{0.04} & 3.94\ci{0.02} & 0.80 & 49.46 & 0.67 & 3.85\ci{0.05} & 3.89\ci{0.02} \\
EmpathyEar~\cite{EmpathyEar} & 0.83 & 48.02 & 0.72 & 3.93\ci{0.01} & 3.95\ci{0.02} & 0.82 & 48.69 & 0.66 & 3.88\ci{0.02} & 3.90\ci{0.03} \\
UniTalker~\cite{UniTalker}  & 0.90 & 42.01 & 0.74 & 4.13\ci{0.02} & 4.10\ci{0.02} & 0.88 & 44.57 & 0.70 & 4.05\ci{0.03} & 4.08\ci{0.03} \\ 
\hhline{-|---|--|---|--}
\rowcolor[HTML]{F5F9FE} 
\textbf{FacialTalker (D)} & \textbf{0.92} & \textbf{39.29} & \textbf{0.79} & \textbf{4.22}\ci{0.02} & \textbf{4.23}\ci{0.04} & \textbf{0.93} & \textbf{40.12} & \textbf{0.76} & \textbf{4.14}\ci{0.02} & \textbf{4.13}\ci{0.01} \\
\rowcolor[HTML]{F5F9FE} 
\textbf{FacialTalker (C)} & 0.91$^{*}$ & 40.11$^{*}$ & 0.78$^{*}$ & 4.19$^{*}$\ci{0.01} & 4.20$^{*}$\ci{0.03} & 0.91$^{*}$ & 42.68$^{*}$ & 0.74$^{*}$ & 4.11$^{*}$\ci{0.03} & \textbf{4.13}\ci{0.02} \\ 
\hhline{-|---|--|---|--}
~~-~FT-CLIP & 0.89 & 41.54 & 0.72 & 4.14\ci{0.02} & 4.13\ci{0.01} & 0.90 & 43.98 & 0.69 & 4.07\ci{0.01} & 4.06\ci{0.01} \\
~~-~FT-base & 0.90 & 41.35 & 0.76 & 4.16\ci{0.01} & 4.15\ci{0.02} & 0.91$^*$ & 43.15 & 0.72 & 4.10\ci{0.02} & 4.11$^{*}$\ci{0.02} \\ 
\hhline{-|---|--|---|--}
\end{tabular}
}
\end{table*}

\begin{table}[t]
\caption{\label{tab:exp-facialtalker-2} Subjective (95\% confidence interval) and objective experimental results on VSDD-1K datasets.}
\vspace{-2mm}
\resizebox{1\linewidth}{!}{
\begin{tabular}{l|ccc|cc}
\hhline{-|---|--}
\multirow{2}{*}{\textbf{Models}}
  & \makecell{\textbf{SIM}$\uparrow$}
  & \makecell{\textbf{PDTW}$\downarrow$}
  & \makecell{\textbf{ACC$_{E}$}$\uparrow$}
  & \multicolumn{1}{c}{\makecell{\textbf{MOS$_{N}$}$\uparrow$}}
  & \multicolumn{1}{c}{\makecell{\textbf{MOS$_{E}$}$\uparrow$}} \\
\cline{2-6}
\hhline{~|---|--}
  & \multicolumn{5}{c}{\cellcolor[HTML]{EFEFEF} \textit{Dataset VSDD-1K}} \\ 
\hhline{-|---|--}
\rowcolor[HTML]{FDF0EF} 
\multicolumn{1}{l|}{Ground Truth} & - & - & - & 4.47\ci{0.02} & 4.36\ci{0.04} \\
\hhline{-|---|--}
GRU-CSS~\cite{GRU-CSS}    & 0.71 & 70.26 & 0.51 & 3.45\ci{0.04} & 3.51\ci{0.03} \\
M$^2$CTTS~\cite{M2-ctts}  & 0.74 & 68.56 & 0.58 & 3.53\ci{0.03} & 3.55\ci{0.02} \\
MSRGCN~\cite{MSRGCN}     & 0.75 & 65.11 & 0.60 & 3.62\ci{0.04} & 3.67\ci{0.04} \\
ECSS~\cite{ECSS}       & 0.77 & 61.45 & 0.62 & 3.68\ci{0.02} & 3.71\ci{0.04} \\
GPT-Talker~\cite{GPT-Talker} & 0.88 & 47.60 & 0.65 & 3.88\ci{0.03} & 3.89\ci{0.01} \\ 
\hhline{-|---|--}
Empatheia~\cite{Empatheia}  & 0.81 & 49.21 & 0.68 & 3.86\ci{0.05} & 3.93\ci{0.03} \\
EmpathyEar~\cite{EmpathyEar} & 0.82 & 49.62 & 0.68 & 3.87\ci{0.02} & 3.95\ci{0.03} \\
UniTalker~\cite{UniTalker}  & 0.91$^{*}$ & 45.38 & 0.71 & 4.08\ci{0.03} & 4.11\ci{0.02} \\ 
\hhline{-|---|--}
\rowcolor[HTML]{F5F9FE} 
\textbf{FacialTalker (D)} & \textbf{0.92} & \textbf{40.10} & \textbf{0.78} & \textbf{4.17}\ci{0.02} & \textbf{4.19}\ci{0.03} \\
\rowcolor[HTML]{F5F9FE} 
\textbf{FacialTalker (C)} & \textbf{0.92} & 41.29$^{*}$ & 0.76$^{*}$ & 4.15$^{*}$\ci{0.04} & 4.17$^{*}$\ci{0.03} \\ 
\hhline{-|---|--}
~~-~FT-CLIP & 0.90 & 43.67 & 0.71 & 4.11\ci{0.01} & 4.09\ci{0.02} \\
~~-~FT-base & \textbf{0.92} & 41.52 & 0.75 & 4.13\ci{0.02} & 4.14\ci{0.01} \\ 
\hhline{-|---|--}
\end{tabular}
}
\end{table}

\subsection{FacialTalker Validity Verification \label{sec-7.3}}
We randomly select 200 dialogue samples from each dataset for evaluation. Tables~\ref{tab:exp-facialtalker-1} and~\ref{tab:exp-facialtalker-2} show that FacialTalker consistently outperforms conventional CSS models and visual-aware baselines on MultiDialog, AvaMERG, and VSDD-1K. In particular, FacialTalker achieves the highest speaker similarity and the lowest prosody distance on nearly all datasets, indicating a strong ability to preserve speaker identity and model natural prosodic patterns.
For example, FacialTalker (D) obtains SIM/PDTW scores of 0.92/39.29 on MultiDialog, 0.93/40.12 on AvaMERG, and 0.92/40.10 on VSDD-1K, consistently surpassing UniTalker. FacialTalker also achieves the best emotional performance, with ACC\(_E\) scores of 0.79, 0.76, and 0.78 on the three datasets, respectively. In subjective evaluation, the results further support these findings: FacialTalker (D) achieves the highest MOS\(_N\) and MOS\(_E\) scores across all datasets and remains closest to the ground truth, demonstrating superior naturalness and emotional appropriateness.
In addition, FT-base and UniTalker mainly differ in their visual tokenizer design. The superior performance of FT-base over UniTalker suggests that AU-based facial expression modeling, compared with 128 facial landmarks, captures richer emotional information from facial cues.
Overall, these results demonstrate that FacialTalker effectively integrates facial expression cues with dialogue context, leading to more natural, expressive, and speaker-consistent speech generation than existing methods.

\subsection{Ablation Study}
We further conduct ablation studies to evaluate the contributions of the AUTokenizer-based visual representation and the DualDPO post-training strategy. As shown in Tables~\ref{tab:exp-facialtalker-1} and~\ref{tab:exp-facialtalker-2}, removing either component consistently degrades performance across all datasets. 
In particular, compared with FT-CLIP, FacialTalker (D) improves \(\mathrm{ACC}_E\) from \(0.72\) to \(0.79\) and \(\mathrm{MOS}_E\) from \(4.13\) to \(4.23\) on MultiDialog, and reduces \(\mathrm{PDTW}\) from \(43.67\) to \(40.10\) on VSDD-1K. Compared with FT-base, it further improves \(\mathrm{ACC}_E\) from \(0.72\) to \(0.76\) on AvaMERG and reduces \(\mathrm{PDTW}\) from \(41.35\) to \(39.29\) on MultiDialog. 
These results demonstrate that both AUTokenizer and DualDPO are critical to the final performance.

\subsection{Visualization Analysis}
\begin{figure}[t]
\centering
\centerline{
\includegraphics[width=1\linewidth]{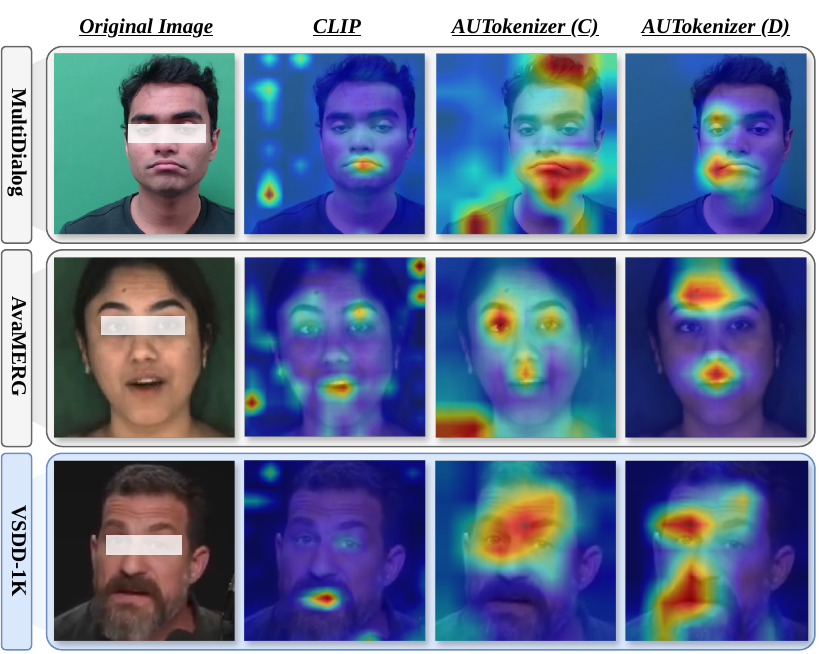}
}
\caption{Visualization of Facial Expression Attention Between CLIP~(text prompt ``facial expression'') and AUTokenizer.}
\label{fig:demo-v2}
\end{figure}

Fig.~\ref{fig:demo-v2} presents visualization results on MultiDialog, AvaMERG, and VSDD-1K. Compared with CLIP, AUTokenizer-based representations attend more consistently to expression-relevant facial regions, such as the eyebrows, eyes, and mouth, rather than to scattered background or weakly related areas. 
Moreover, AUTokenizer (D) produces more compact and discriminative activations over these regions than AUTokenizer (C), likely due to the larger amount of facial data in DISFA, which improves the robustness of the learned visual representations. This observation is consistent with the results reported in Section~\ref{sec-7.3}. 
Overall, the qualitative results align well with the quantitative findings and further highlight the importance of AUTokenizer in enabling FacialTalker to generate expressive speech.

\section{Conclusion}
In this work, we present FacialTalker, which enables strong facial affect perception and effective multimodal contextual modeling with LLMs. We design AUTokenizer to encode rich facial expression cues into discrete tokens by predicting combinations of facial Action Units with an FSQ-based quantization layer. AUTokenizer compresses each facial frame into a single token, providing a more natural interface for LLM-based modeling while significantly reducing training and inference costs.
We further propose a dual direct preference optimization strategy for bimodal contextual understanding. This post-training scheme encourages the model to jointly capture users' facial expressions and speech semantics. In addition, we construct a large-scale visual--speech dialogue dataset through an automated pipeline, comprising approximately 1,033 hours of conversational speech and more than 85\% valid facial frames. Extensive experiments demonstrate the effectiveness of the proposed dataset and further show that incorporating facial affect cues improves contextual understanding, leading to more natural, expressive, and contextually appropriate speech generation.


\clearpage
\section{Acknowledgments}
This research of Rui Liu was funded by the General Program (No. 62476146)
of the National Natural Science Foundation of China, the Young Elite
Scientists Sponsorship Program by CAST (No. 2024QNRC001), the Outstanding
Youth Project of Inner Mongolia Natural Science Foundation (No. 2025JQ011),
the Key R\&D and Achievement Transformation Program of Inner Mongolia
Autonomous Region (No. 2025YFHH0014), and the Central Government Fund for
Promoting Local Scientific and Technological Development (No. 2025ZY0143). The work of Haizhou Li was supported by the Shenzhen Science and Technology
Research Fund (Fundamental Research Key Project, Grant No.
JCYJ20220818103001002) and the Program for Guangdong Introducing Innovative
and Entrepreneurial Teams (Grant No. 2023ZT10X044).


\bibliographystyle{ACM-Reference-Format}
\balance
\bibliography{acmart}

\end{document}